\documentstyle[12pt]{article}
\begin{document}
\textwidth 19 true cm
\textheight 23 true cm
\baselineskip=12pt
\voffset=-32 mm


\renewcommand{\baselinestretch}{1.5}
\renewcommand{\theequation}{\arabic{equation}}
\def\brho{\mbox{\boldmath $\rho $}}


\title{{Orientational Ordering of Polymers}
\\{on a Fluctuating Flexible Surface}}
\author{ R.Podgornik\thanks{On leave from J.Stefan Institute, P.O.B. 100,
Ljubljana, Slovenia.}
\\{Laboratory of Structural Biology}
\\{Division of Computer Research and Technology}
\\{National Institutes of Health, Bethesda, MD 20892} }

\date{}
\begin{titlepage}
\maketitle
\baselineskip=10pt
\begin{abstract}
Arguments are presented to the effect that embedding semi-flexible
(wormlike) ideal polymers into a fluctuating, flexible surface leads
to an effective attractive orientational interaction between polymer
segments that precipitates an orientational ordering transition of the
polymer chains on the surface even in the case of otherwise ideal
(non-interacting) chains. The orientational interaction is analogous
to the (zero order) Casimir force and is due to the energy change in
surface conformational fluctuations in the presence of embedded
semi-flexible chains.
\centerline{\today }
\end{abstract}
\end{titlepage}


\textwidth 19 true cm
\textheight 23 true cm
\baselineskip=15pt
\parskip=15pt plus 1pt
\parindent=40pt


\section{Introduction}

Polymers or polymer-like structures confined to lye in an embedding
surface can appear in the context of different membrane and/or interfacial
phenomena. Partially polymerized membranes of unsaturated amphiphiles
can sometimes make long trains of joined monomers that behave as
surface - embedded polymer chains \cite{bib1}. Hydrophobic polymers in
aqueous solutions can become embedded or multiply attached to
hydrocarbon interior of membranes \cite{bib1}, while certain polymers
can also form monomolecular films on the water-air interface
\cite{bib2}. It has been observed that embedded polymers in the case
of partially polymerized membranes can promote wrinkling transition in
membranes \cite{bib3} while multiply anchored polymers give rise to
bulging and budding of membranes \cite{bib4}. In the case of interface
polymer films \cite{bib2} it has been shown that they make segregated
domains even at very low (submonolayer) concentrations where one would
expect the polymers to be evenly distributed over the surface.

This plethora of new phenomena that appear closely connected with the
fact that polymers are not only adsorbed but actually embedded into a
supporting surface have recently become a focus of several theoretical
studies \cite{bib5}. In this contribution we try to find out what is
the influence of finite temperature, which in general leads to shape
fluctuations in flexible (membrane) surfaces or interfaces, on the
properties of embedded polymer chains. In the framework of a
semiflexible polymer chain model we show that finite temperature leads
to membrane fluctuation generated effective orienting forces between
different polymer segments. This interaction formally resembles the
usual nematic interaction used in various models of the nematic
ordering of long flexible chains \cite{bib9} and is thus not
surprising that it leads to an orientational ordering transition.

The interesting point is that this transition is driven by shape
fluctuations of the supporting surface into which the polymer is
embedded.

The problem we address here is related to the recent investigations of
fluctuation-induced forces between manifolds immersed in a correlated
fluid \cite{kardar}. The existence of surface fluctuations promoted
attraction between different segments of the embedded polymer chain
can also be related to the zero - order Casimir forces (Keesom forces)
between beads attached to flexible membranes \cite{bib8}. Other
related problems include adsorption of polymers on soft, fluctuating
surfaces \cite{bib6} or the collapse of polymers on fluctuating
surfaces \cite{bib7}.

The outline of the paper is as follows: I shall first derive the
partition function of a surface embedded semiflexible polymer chain at
zero temperature. Then I shall relax the zero temperature constraint
and derive the effective partition function for a polymer, after the
supporting surface degrees of freedom have been integrated out. At
last an approximate solution for this effective partition function will be
given and discussed.

\section{Analysis}
\subsection{Semiflexible polymer chain embedded in a surface}

I use a semiflexible (worm - like chain) model to describe the polymer
chain. Though the theory presented below is formulated for a single
long polymer chain, it can be trivially generalized to a many - chain
system by assuming that the variable $n$ has a continuous part $N_{\alpha}$,
describing a single chain, and a discrete part describing different
chains $\alpha$.

I investigate the statistical properties of a semiflexible polymer chain
embedded in a fluctuating surface. Let us assume that the n-th bead of
the polymer has coordinates ${\bf r}(n) = (x(n),y(n),z(n))$ and that
the quenched profile of the embedding surface can be written in a
Monge parametrization $z = \zeta(x,y) = \zeta(\brho )$, where I introduced the
two dimensional vector $\brho = (x,y)$. The embedding {\sl ansatz} for
the polymer can thus be written in a form
\begin{equation}
z(n) = \zeta(\brho (n)).
\label{equ1}
\end{equation}
I also assume that the polymers can be described in terms of the
semiflexible chain model, where the configurational energy is given as
\begin{equation}
\beta{\cal H}_0({\bf r}(n)) = {\textstyle\frac{1}{2}}\beta\epsilon \int_{0}^{N}
\left(\frac{d^2{\bf r}(n)}{dn^2}\right)^2~dn =
{\textstyle\frac{1}{2}}\beta\epsilon \int_{0}^{N} \left(\ddot{\bf
r}(n)\right)^2~dn
\label{equ2}
\end{equation}
where $\beta\epsilon = {\textstyle\frac{1}{\ell^2}}$, with $\ell$
being the persistence length and $N$ is the number of monomers. For a
differentiable curve the tangent $\dot{\bf r}(n)$ is a unit vector,
$\dot{\bf r}(n)^2 = 1$. The partition function of a semiflexible chain
embedded in the surface $\zeta(x,y)$ can thus be written in a succinct
form
\begin{equation}
\Xi\left(\zeta \left(\brho (n)\right)\right) = \int\!\! \dots \!\!\int {\cal
D}{\bf r}(n) \prod_{n}\delta\left(z(n) - \zeta(\brho (n))\right)
\prod_{n}\delta\left( \dot{\bf r}(n)^2 - 1 \right) \times
e^{-\beta{\cal H}_0({\bf r}(n))}.
\label{equ4}
\end{equation}
The explicit incorporation of the constraint that the local tangent
has to be a unit vector permits us to perform the functional
integration in Eq.\ref{equ4} over an unconstrained set of ${\bf r}(n)$.

By integrating out the $z(n)$ variables and introducing an integral
representation for the delta function I obtain the partition function
in the form
\begin{equation}
\Xi\left(\zeta \left(\brho (n)\right)\right) = \int\!\! \dots \!\!\int{\cal
D}{\brho}(n) {\cal D}\lambda (n)~e^{-\beta{\cal H}({\brho}(n),\zeta(\rho (n)))}
\label{equ5}
\end{equation}
where the effective configurational Hamiltonian reads
\begin{equation}
\beta{\cal H}({\brho}(n),\zeta(\brho (n))) =
{\textstyle\frac{1}{2}}\beta\epsilon \int_{0}^{N}
\left( \ddot{\brho}(n)^2 + \ddot{\zeta}(\brho(n))^2\right)dn~ +
{}~\imath\!\!\int_{0}^{N}\!\!\lambda (n)
\left(\dot{\brho}(n)^2 + \dot{\zeta}(\brho(n))^2\right)dn~.
\label{equ6}
\end{equation}
It is at this point where a first major approximation is necessary in
order to proceed. Instead of enforcing the condition $\dot{\bf r}(n)^2
= 1 $ locally at each point along the polymer chain, we presume that
it is only satisfied globally \cite{glob} {\sl i.e.} on the average
\begin{equation}
\left< \dot{\bf r}(n)^2 \right> = \left< \dot{\brho}(n)^2 +
\dot{\zeta}\left(\brho(n)\right)^2 \right> = 1,
\label{equ9}
\end{equation}
where the average is over all polymer configurations. Thus the
magnitude of the tangent can fluctuate locally but on the average,
over the entire polymer chain, it is fixed. Replacement of the local
continuity constraint with a global one Eq.~\ref{equ9} can also
expressed as $\lambda (n) \longrightarrow \lambda = const.$. Thus one
remains (after making at the same time also the transformation $\imath\lambda
\longrightarrow \lambda$) with the following form of the partition function
\begin{equation}
\Xi\left(\zeta \left(\brho (n)\right)\right) = \int\!\! \dots
\!\!\int{\cal D}\brho(n)~exp\left(- \beta{\cal H}(\brho (n)) \right),
\label{equ12}
\end{equation}
where some irrelevant constant terms have been omitted. Also
\begin{eqnarray}
\beta{\cal H}(\brho (n)) &=&
{\textstyle\frac{1}{2}}\beta\epsilon \int_{0}^{N} \left(
\ddot{\brho}(n)^2 + \ddot{\zeta}(n)^2 \right)dn~ + \nonumber\\
&+& ~\lambda\!\!\int_{0}^{N}\!\!  \left(\dot{\brho}(n)^2 +
\dot{\zeta}(n)^2 \right)dn~  \nonumber\\
&\cong& {\textstyle\frac{1}{2}}\beta\epsilon \int_{0}^{N} \left(
\ddot{\brho}(n)^2 + \frac{\partial \zeta(\brho )}{\partial
\rho_i}\frac{\partial \zeta(\brho )}{\partial \rho_k}
\ddot{\brho}_i(n)\ddot{\brho}_k(n) \right)dn~ + \nonumber\\
&+& ~\lambda\!\!\int_{0}^{N}\!\!  \left(\dot{\brho}(n)^2 + \frac{\partial
\zeta(\brho )}{\partial
\rho_i}\frac{\partial \zeta(\brho )}{\partial \rho_k}
\dot{\brho}_i(n)\dot{\brho}_k(n) \right)dn~ + \dots , \nonumber\\
{}~
\label{equ13}
\end{eqnarray}
where we have limited ourselves to the terms of the second order in
the derivatives of ${\brho}(n)$.

\subsection{Conformational fluctuations of the embedding surface}

Going now to a non-zero temperature one has to take into account that
the embedding surface is allowed to fluctuate, taking the polymer with
it on its conformational wanderings. The unrestrained partition
function in this case is
\begin{equation}
\Xi\left( N \right) = \int\!\! \dots
\!\!\int{\cal D}\brho(n) {\cal D}\zeta \left(\brho\right)~exp\left(
-\beta{\cal H}_T\left( \zeta \left(\brho\right), \brho (n)\right)
\right)
\label{equ14}
\end{equation}
where after introducing the Fourier transform $\zeta(\brho) = \sum_{\bf Q}
\zeta({\bf Q})~e^{\imath {\bf Q}\brho}$
\begin{eqnarray}
\beta{\cal H}_T\left( \zeta \left(\brho\right), \brho (n)\right) &=&
{\textstyle\frac{1}{2}}\beta\epsilon \int_{0}^{N}
\ddot{\brho}(n)^2dn~ + ~\lambda\!\!\int_{0}^{N}\!\!
\dot{\brho}(n)^2dn~ + \nonumber\\
& &{\textstyle\frac{1}{2}}\sum_{\bf Q,Q'}
\zeta({\bf Q}){\cal D}({\bf Q},{\bf Q}')\zeta(-{\bf Q}')~ +
{}~{\textstyle\frac{1}{2}} \beta \sum_{\bf Q} V\left(\zeta({\bf Q}) \right)
\nonumber\\
{}~
\label{equ15}
\end{eqnarray}
where
\begin{eqnarray}
{\cal D}({\bf Q},{\bf Q}';\brho (n)) &=& \beta\epsilon~\int_0^N dn~\left( {\bf
Q}\ddot{\brho}(n) \right) \left( {\bf Q}'\ddot{\brho}(n)
\right)~e^{\imath\left( {\bf Q}-{\bf Q}'\right)\brho(n)}~ + \nonumber\\
&+&~ 2\lambda~~\int_0^N dn~\left( {\bf
Q}\dot{\brho}(n) \right) \left( {\bf Q}'\dot{\brho}(n)
\right)~e^{\imath\left( {\bf Q}-{\bf Q}'\right)\brho(n)}.
\label{equ16}
\end{eqnarray}
$V\left(\zeta({\bf Q}) \right)$ is the intrinsic deformation energy of
the surface describing the energetics of the surface in the absence of
embedded polymers, I assume it to be of the form
\begin{equation}
V\left(\zeta({\bf Q}) \right) = V(Q)~\vert\zeta({\bf Q}) \vert^2.
\label{equ17}
\end{equation}
With this {\sl ansatz} one can integrate out the $\zeta (\brho )$
degrees of freedom obtaining
\begin{eqnarray}
\kern-50pt\Xi\left(\brho (n) \right) &=& \int\!\!\dots\!\!\int{\cal
D}\zeta\left(\brho\right)~\Xi\left(\zeta \left(\brho\right), \brho (n)
\right) = \nonumber\\
&= & \int\!\!\dots\!\!\int{\cal D}\brho(n)
\times exp\left( -{\textstyle\frac{1}{2}}\beta\epsilon \int_{0}^{N}
\ddot{\brho}(n)^2dn~ - ~\lambda\!\!\int_{0}^{N}\!\!
\dot{\brho}(n)^2dn~ - ~{\textstyle\frac{1}{2}}Tr~\ln{{\cal G}({\bf
Q},{\bf Q}';\brho (n))}  \right) \nonumber\\
{}~
\label{equ18}
\end{eqnarray}
with
\begin{equation}
{\cal G}({\bf Q},{\bf Q}';\brho (n)) = \left(\beta V(Q)\right)~\delta({\bf
Q}-{\bf Q}') +
{\cal D}({\bf Q},{\bf Q}';\brho (n)).
\label{equ19}
\end{equation}
At this point I use the standard identity relating $Tr~\ln$ of an
operator with a coupling integral of its resolvent
\begin{equation}
Tr~\ln{{\cal G}({\bf Q},{\bf Q}';\brho (n))} = Tr~\int_0^1~d\mu {\cal
R}_{\mu}({\bf Q},{\bf Q}';\brho (n)),
\label{equ20}
\end{equation}
where the resolvent ${\cal R}_{\mu}({\bf Q},{\bf Q}';\brho (n)) $ is
defined as
\begin{equation}
{\cal R}_{\mu}({\bf Q},{\bf Q}';\brho (n)) = {\cal D}({\bf Q},{\bf
Q}';\brho (n))~\left(\beta V(Q)\right)^{-1}\times\left( {\cal I} + \mu{\cal
D}({\bf Q},{\bf
Q}';\brho (n))~\left(\beta V(Q)\right)^{-1} \right)^{-1},
\label{equ21}
\end{equation}
with $\cal I$ being the identity operator in $Q$ space. I have omitted
a term that does not depend on the coordinates $\brho(n)$ from
Eq.\ref{equ20}. Evaluating the resolvent to the first order in $\mu$ I
obtain
\begin{equation}
Tr~\ln{{\cal G}({\bf Q},{\bf Q}';\brho (n))} \cong Tr~{\cal D}({\bf Q},{\bf
Q}';\brho (n))~\left(\beta V(Q)\right)^{-1} -
{\textstyle\frac{1}{2}}~Tr~\left({\cal
D}({\bf Q},{\bf Q}';\brho (n))~\left(\beta V(Q)\right)^{-1}\right)^2 + \dots.
\label{equ22}
\end{equation}
What one remains with at the end is an effective interaction between
polymer segments as if they would be confined to a planar surface. It
has the form
\begin{eqnarray}
\kern-70pt{\textstyle\frac{1}{2}}Tr~\ln{{\cal G}({\bf Q},{\bf
Q}';\brho (n))} &\cong&  \frac{\beta\epsilon}{2} \int_0^N dn \ddot{\brho}_i(n)
\ddot{\brho}_k(n) \sum_{\bf Q} \frac{{\bf Q}_i{\bf Q}_k}{\left(\beta
V(Q)\right)}~ + \nonumber\\
&+& \lambda \int_0^N dn \dot{\brho}_i(n)
\dot{\brho}_k(n)\frac{{\bf Q}_i{\bf Q}_k}{\left(\beta V(Q)\right)}~ - \nonumber
\end{eqnarray}
\begin{equation}
\kern-70pt - ~\frac{(2\lambda)^2}{4} \int_0^N\!\!\int_0^N dn dn'
\dot{\brho}_i(n)
\dot{\brho}_k(n) \dot{\brho}_l(n')\dot{\brho}_m(n')~ \sum_{\bf Q}
\frac{{\bf Q}_i{\bf Q}_k}{\left(\beta V(Q)\right)}~e^{\imath{\bf Q}(\brho (n) -
\brho(n'))}~
\sum_{\bf Q'}
\frac{{\bf Q'}_l{\bf Q'}_m}{\left(\beta V(Q')\right)}~e^{-\imath{\bf Q'}(\brho
(n) - \brho(n'))}~~
+ \dots \nonumber\\
{}~
\label{equ23}
\end{equation}
I have limited myself to second order terms in the derivatives of $\brho
(n)$ at any $n$ (self - energy term) and to fourth order terms for two
different values of $n$ (interaction term).

Introducing now the orientational correlation function between the
directions of the surface normals ${\bf n}(\brho)$ in the absence of
polymer chains and for small surface deformations as
\begin{equation}
\left< {\bf n}_i(\brho){\bf n}_k(\brho') \right> \cong
\left< \frac{\partial \zeta(\brho)}{\partial \brho_i}\frac{\partial
\zeta(\brho')}{\partial \brho'_k} \right> =
 - \sum_{\bf Q,Q'}
{\bf Q}_i{\bf Q}'_k \left< \zeta({\bf Q}) \zeta({\bf
Q}')\right>~e^{\imath ({\bf Q}\brho + {\bf Q}'\brho')}
\label{equ24}
\end{equation}
where due to the harmonic {\sl ansatz} for the configurational energy
of the bare surface I have
\begin{equation}
\left<\zeta({\bf Q})\zeta({\bf Q}')\right> =
{\textstyle\frac{1}{2}}\frac{kT}{V(Q)}~\delta(
{\bf Q} + {\bf Q}'),
\label{equ25}
\end{equation}
thus
\begin{equation}
\beta \left< {\bf n}_i(\brho){\bf n}_k(\brho') \right> =
{\textstyle\frac{1}{2}}\beta{\cal
F}_{ik}(\brho - \brho') = {\textstyle\frac{1}{2}}~\sum_{\bf Q}
\frac{{\bf Q}_i{\bf Q}_k}{V(Q)}~e^{\imath{\bf Q}(\brho - \brho')}.
\end{equation}
The final expression for the partition function of a surface -
embedded polymer chain at finite temperature is thus
\begin{equation}
\Xi\left( N \right) = \int\!\!\dots\!\!\int{\cal
D}\brho(n)~e^{-\beta{\cal
H}_{eff}(\dot{\brho}(n),\ddot{\brho}(n))}
\label{equ26}
\end{equation}
with
\begin{eqnarray}
\kern-50pt\beta{\cal H}_{eff}(\dot{\brho}(n),\ddot{\brho}(n)) &=&
{\textstyle\frac{1}{2}}\beta\epsilon \int_{0}^{N}
\ddot{\brho}(n)^2dn~ + ~\lambda\!\!\int_{0}^{N}\!\!
\dot{\brho}(n)^2dn~ + \nonumber\\
&+& \beta\epsilon \int_0^N dn \ddot{\brho}_i(n)
\ddot{\brho}_k(n) \left< {\bf n}_i(0){\bf n}_k(0)
\right> + \nonumber\\
&+& 2\lambda \int_0^N dn \dot{\brho}_i(n)
\dot{\brho}_k(n) \left< {\bf n}_i(0){\bf n}_k(0) \right> - \nonumber\\
&-& 4\lambda^2 \int_0^N\!\!\int_0^N dn dn' \dot{\brho}_i(n)
\dot{\brho}_k(n) \dot{\brho}_l(n')\dot{\brho}_m(n') \left< {\bf
n}_i(n){\bf n}_m(n') \right> \left< {\bf n}_k(n'){\bf n}_l(n) \right>
+ \dots \nonumber\\
{}~
\label{equ27}
\end{eqnarray}
Obviously the integration of the surface degrees of freedom leads to
an effective {\sl attraction} between the polymer segments, which is
orientation dependent. This attraction has the same origin as the
usual (zero order) Casimir force \cite{bib11} and we could simply dubb
it the nematic Casimir force, or maybe more appropriately the nematic
Keesom force. The fluctuations of the embedding surface play in this
case the role of vacuum fluctuations of the electromagnetic field
\cite{bib10}.

\subsection{Effective embedded - polymer partition function and
orientational ordering}

Introducing now the orientational tensor $\sigma_{ik}(\brho)$ as \cite{doi}
\begin{equation}
\sigma_{ik}({\brho}) = \int_0^N dn ~\dot{\brho}_i(n)\dot{\brho}_k(n)
{}~\delta({\brho} - {\brho}(n))
\label{equ28}
\end{equation}
I can write the partition function alternatively as
\begin{equation}
\kern-50pt\Xi\left( N \right) = \int\!\!\dots\!\!\int{\cal
D}\brho(n){\cal D}\sigma_{ik}({\brho})~ \delta\!\!\left( \int_0^N\!\!\!dn
{}~\dot{\brho}_i(n)\dot{\brho}_k(n)~
\delta({\brho}-{\brho}(n)) -  \sigma_{ik}({\brho})\right)~e^{-\beta{\cal
H}_{eff}(\dot{\brho}(n),\ddot{\brho}(n),\sigma_{ik}(\brho))},
\label{equ29}
\end{equation}
where
\begin{eqnarray}
\kern-60pt\beta{\cal
H}_{eff}(\dot{\brho}(n),\ddot{\brho}(n),\sigma_{ik}(\brho)) &=&
{\textstyle\frac{1}{2}}\beta\epsilon \int_{0}^{N}
\ddot{\brho}(n)^2dn~ + ~\lambda\!\!\int_{0}^{N}\!\!
\dot{\brho}(n)^2dn~+ \nonumber\\
&+&  {\textstyle\frac{1}{2}}\beta\epsilon \int_0^N dn ~\ddot{\brho}_i(n)
\ddot{\brho}_k(n)~ {\cal F}_{ik}(0) + \nonumber\\
&+& \lambda \int d^2\brho~ \sigma_{ik}({\brho})~{\cal F}_{ik}(0)
- \nonumber\\
&-& \lambda^2 \int\!\!\int d^2\brho
d^2\brho'~\sigma_{ik}({\brho})\sigma_{lm}({\brho}')~ {\cal
F}_{im}({\brho}-{\brho}'){\cal F}_{kl}({\brho}-{\brho}') + \dots
\nonumber\\
{}~
\label{equ30}
\end{eqnarray}
In order to proceed from here one has to specify the form of $V(Q)$,
valid for the bare surface. To get some general trends and estimates I
shall limit myself to the following simple form of ${\cal
F}_{ik}(\brho)$
\begin{equation}
{\cal F}_{ik}(\brho) = \delta_{ik}~F(\vert \brho \vert).
\label{equ31}
\end{equation}
Also writing the delta function in  an integral representation
I obtain the following form of the effective Hamiltonian
\begin{eqnarray}
\kern-60pt\beta\hat{\cal H}\left(\dot{\brho}(n),\ddot{\brho}(n),
\sigma_{ik}(\brho),\psi_{ik}(\brho)\right) &=& \beta{\cal
H}_{eff}(\dot{\brho}(n),\ddot{\brho}(n),\sigma_{ik}(\brho))~ + ~\imath
\int_0^N dn ~\dot{\brho}_i(n)\dot{\brho}_k(n)~\psi_{ik}(\brho(n)) - \nonumber\\
&-& \imath \int d^2\brho~\psi_{ik}(\brho)\sigma_{ik}({\brho})~ = \nonumber\\
&=& {\textstyle\frac{1}{2}}\beta\epsilon\left(
1 + F(0)\right)~\int_{0}^{N} \ddot{\brho}(n)^2dn~ +
{}~\lambda\!\!\int_{0}^{N}\!\! \dot{\brho}(n)^2dn~ + \nonumber\\
&+& ~(\lambda F(0))
\int d^2\brho~ Tr(\sigma_{ik}({\brho}))~ - \nonumber\\
&-& ~\lambda^2
\int\!\!\int d^2\brho
d^2\brho'~Tr\left(\sigma_{ik}({\brho})\sigma_{kl}({\brho}')\right)F^2(\vert
\brho - \brho'\vert) ~ +\nonumber\\
&+& \imath \int_0^N dn
{}~\dot{\brho}_i(n)\dot{\brho}_k(n)~\psi_{ik}(\brho(n)) - \imath \int
d^2\brho~\psi_{ik}(\brho)\sigma_{ik}({\brho}) \nonumber\\
{}~
\label{equ32}
\end{eqnarray}
and
\begin{equation}
\Xi\left( N \right) = \int\!\!\dots\!\!\int{\cal
D}\brho(n){\cal D}\sigma_{ik}({\brho}){\cal D}\psi_{ik}(\brho)~exp\left(
-\beta\hat{\cal H}(
\dot{\brho}(n),\ddot{\brho}(n),\sigma_{ik}(\brho),\psi_{ik}(\brho))\right) .
\label{equ33}
\end{equation}
The above partition function can not be evaluated explicitely because
$\psi_{ik}(\brho(n))$ depends on the polymer coordinates and the
functional integrals with respect to $\brho(n)$ and
$\psi_{ik}(\brho(n))$ can not be separated. However, as recognized by
Gupta and Edwards \cite{bib9} another approximation suggests
itself. Replacing the fields $\sigma_{ik}(\brho) $ and
$\psi_{ik}(\brho) $ by their average values $\bar{\sigma}_{ik} $ and
$\bar{\psi}_{ik} $, independent of $\brho$ ( and at the same time
making the transformation $\imath\bar{\psi}_{ik} \longrightarrow
\bar{\psi}_{ik} $), one can first of all explicitely evaluate the
$\brho(n)$ part of the functional integral
\begin{eqnarray}
\kern-70pt\int\!\!\dots\!\!\int&{\cal D}\brho(n)&\!\!e^{-
{\textstyle\frac{1}{2}}\beta\epsilon\left( 1 +
F(0)\right)~\int_{0}^{N} \ddot{\brho}(n)^2dn~ -
{}~\lambda\!\!\int_{0}^{N}\!\! \dot{\brho}(n)^2dn~ - ~\int_0^N dn
{}~\dot{\brho}_i(n)\dot{\brho}_k(n)~\bar{\psi}_{ik}} =
\nonumber\\
&=& ~\int\!\!\dots\!\!\int{\cal D}\brho(n)~e^{ -2\pi
{}~\sum_{\jmath=-\infty}^{\jmath=+\infty} \brho_i(\jmath) {\cal
M}_{ik}(\jmath) \brho_k(-\jmath)}~ = ~e^{ -{\textstyle\frac{1}{2}}
{}~\sum_{\jmath=-\infty}^{\jmath=+\infty} \ln{Det {\cal
M}_{ik}(\jmath)}}\nonumber\\
{}~
\label{equ34}
\end{eqnarray}
where
\begin{equation}
{\cal M}_{ik}(\jmath) =
\left({\textstyle\frac{1}{2}}\beta\epsilon\left( 1 +
F(0)\right)\left(\textstyle\frac{2\pi \jmath}{N}\right)^4~ + ~\lambda
\left(\textstyle\frac{2\pi \jmath}{N}\right)^2 \right)\delta_{ik}~ +
{}~\left(\textstyle\frac{2\pi \jmath}{N}\right)^2 \bar{\psi}_{ik}.
\label{equ35}
\end{equation}
Above I introduced the Rouse modes of the polymer chain as
\begin{equation}
\brho (n) = \sum_{\jmath=-\infty}^{\jmath=+\infty} \brho
(\jmath)~e^{2\pi \imath \jmath \frac{n}{N}}.
\label{equ36}
\end{equation}
The sum of $\ln{Det {\cal M}_{ik}(\jmath)} $ can be evaluated
explicitely and in the limit of long chains where the sum over
$\jmath$ can be replaced with an appropriate integration one obtains
\cite{bib9}
\begin{equation}
{\textstyle\frac{1}{2}}~\sum_{\jmath=-\infty}^{\jmath=+\infty} \ln{Det {\cal
M}_{ik}(\jmath)}~ =
{}~{\textstyle\frac{N}{2}}~\sum_{\alpha}\sqrt{\frac{\left(
\bar\psi_{\alpha} + \lambda \right)}{\frac{\beta\epsilon}{2}\left( 1 +
F(0)\right) }},
\label{equ37}
\end{equation}
where $\alpha$ is the index of the eigenvalue of $\bar\psi_{ik}$. I
have omitted a divergent contribution to the above expression since it
is irrelevant for subsequent developements. In the limit of 'mean
fields' one can now write the free energy corresponding to the
partition function Eq.\ref{equ33} as
\begin{equation}
\kern-30pt\beta{\cal F}_{MF} =
{\textstyle\frac{N}{2}}~\sum_{\alpha}\sqrt{\frac{\left(
\bar\psi_{\alpha}~ + ~\lambda \right)}{\frac{\beta\epsilon}{2}\left( 1 +
F(0)\right) }}~ - ~N\lambda~ - ~{\cal
S}\sum_{\alpha} \bar{\sigma}_{\alpha}\bar{\psi}_{\alpha}~ +
{}~\lambda F(0){\cal S} \sum_{\alpha} \bar{\sigma}_{\alpha}~ -
{}~\lambda^2{\cal S}a_2  \sum_{\alpha} \bar{\sigma}^2_{\alpha}~ +
\dots \nonumber\\
{}~
\label{equ38}
\end{equation}
where ${\cal S} = \int d^2\brho$ is the projected area of the surface
and I introduced the second virial coefficient $a_2$ as
\begin{equation}
a_2 = \int d^2\brho F^2(\vert \brho \vert).
\label{equ39}
\end{equation}
Since the mean - field free energy is a function of $\beta{\cal
F}_{MF} = \beta{\cal F}_{MF}(\lambda, \bar{\sigma}_{\alpha},
\bar{\psi}_{\alpha})$ I get the equilibrium solution by minimizing it
with respect to all three variables:
\begin{eqnarray}
& & \frac{\partial(\beta{\cal F}_{MF})}{\partial\bar{\psi}_{\alpha}} =
0 ~ \Rightarrow ~ \bar{\sigma}_{\alpha} = \frac{{N\over {\cal
S}}}{4\sqrt{{\frac{\beta\epsilon}{2}}{\left( 1 + F(0)\right)}}}~
\frac{1}{\sqrt{\bar{\psi}_{\alpha} + \lambda}}  \nonumber\\
& & \frac{\partial(\beta{\cal F}_{MF})}{\partial\bar{\sigma}_{\alpha}} =
0 ~ \Rightarrow ~ \bar{\psi}_{\alpha} = \lambda F(0) -
2\lambda^2a_2~\bar{\sigma}_{\alpha} + {\cal
O}( \bar{\sigma}^2_{\alpha} ) \nonumber\\
& &  \frac{\partial(\beta{\cal F}_{MF})}{\partial\lambda} =
0 ~ \Rightarrow ~ \sum_{\alpha}\frac{{N\over {\cal
S}}}{4\sqrt{{\frac{\beta\epsilon}{2}}{\left( 1 + F(0)\right)}}}~
\frac{1}{\sqrt{\bar{\psi}_{\alpha} + \lambda}} = {N\over {\cal S}}~ -
F(0)\sum_{\alpha}\bar{\sigma}_{\alpha} + {\cal
O}(\bar{\sigma}^2_{\alpha}).
\nonumber\\
{}~
\label{equ40}
\end{eqnarray}
In order to be consistent the above equations should be linear in the
field $\bar{\sigma}_{\alpha}$.

These equations look very much like the analogous equations obtained
by Gupta and Edwards \cite{bib9} for athermal flexible polymer chains with
orientation dependent attraction. Solving these equations explicitely
for the three fields I obtain the following results. First of all
$\bar{\sigma}_{\alpha}$ is a solution of
\begin{equation}
\left( \bar{\sigma}_{x} - \bar{\sigma}_{y} \right) \left( 2a_2\lambda^2 -
{\frac{\left({\frac{N}{\cal S}}\right)^2}{8(\beta\epsilon)\left( 1 +
F(0)\right)}}  {\frac{(\bar{\sigma}_{x} +
\bar{\sigma}_{y})}{\bar{\sigma}^2_{x}\bar{\sigma}^2_{y}}}~\right).
\label{equ40a}
\end{equation}
{}From the last equation I obtain
\begin{equation}
\sum_{\alpha}\bar{\sigma}_{\alpha} = \frac{{N\over {\cal
S}}}{\left( 1 + F(0)\right)}~
\label{equ40b}
\end{equation}
which is simply a consequence of the continuity of the chain since
\begin{eqnarray}
Tr~\bar{\sigma}_{\alpha} &=& {N\over {\cal S}}~\left< \dot{\brho}^2(n)
\right> = {N\over {\cal S}}~\left( 1 - \left< \dot{z}^2(n)\right>\right) =
\nonumber\\
&=& {N\over {\cal S}}~\left( 1 - \left< \frac{\partial \zeta
(\brho)}{\partial \brho_i}\frac{\partial \zeta
(\brho)}{\partial \brho_k} \dot{\brho}_i(n)\dot{\brho}_k(n)
\right>\right) =  {N\over {\cal S}}~\left( 1 - F(0)\left< \dot{\brho}^2(n)
\right> \right). \nonumber\\
{}~
\label{equ41}
\end{eqnarray}
These are the two equations that have to be solved consistently in
order to get the dependence of the eigenvalues of the orientation
tensor on the surface monomer density, $ \frac{N}{\cal S}$. There is
also a boundary condition that we have to take into account, {\sl
viz.} that in the case of a stiff surface $F \longrightarrow 0$ the
polymer statistics should reduce to the case of a semiflexible chain
embedded in a 2-D surface.

The appropriate solution for $\lambda$ can now be derived from
the first two equations of Eqs.\ref{equ40} in the form
\begin{equation}
\lambda = \frac{\left( 1 + F(0)\right)^2}{2(\beta a_2)
\frac{N}{\cal S}}\left[ 1 - \sqrt{1 - {\frac{(\beta a_2) \left(\frac{N}{\cal
S}\right)^3}{4\epsilon\left( 1
+ F(0)\right)^4}\sum_{\alpha}\frac{1}{\bar{\sigma}^2_{\alpha}} }
}\right] .
\label{equ42}
\end{equation}
The nonisotropic (${\sl i.e.}~\bar{\sigma}_{x} \neq \bar{\sigma}_{y}$)
solution for the orientational tensor can be obtained from
\begin{equation}
{\frac{\bar{\sigma}^2_{x}\bar{\sigma}^2_{y}}{(\bar{\sigma}_{x} +
\bar{\sigma}_{y})}} =
{\frac{\left({\frac{N}{\cal S}}\right)^2}{16(\beta\epsilon)\left( 1 +
F(0)\right)a_2\lambda^2}}.
\label{equ43}
\end{equation}
Defining the orientational order parameter $S$ as
\begin{equation}
\bar{\sigma}_{x,y} = {\frac{\left(N\over {\cal S}\right)}{\left(
1 + F(0)\right)}}~\bigl( 1 \pm S\bigr)
\label{equ44}
\end{equation}
so that $Tr~\bar{\sigma}_{\alpha}$ satisfies the constraint
Eq.~\ref{equ40b}, one can derive an approximate equation for $S$ that has two
solutions. One corresponding to the isotropic state ($S = 0$) and the other one
corresponding to the orientationally ordered phase of polymers ($S \neq 0$),
\begin{equation}
S \cong \Biggl\{ \Biggr.
\begin{array}{ll}
0 & \mbox{;~${\frac{\bigl(1 + F(0)\bigr)^2}{(\beta\epsilon)\lambda^2~a_2
{\frac{N}{\cal S}}}} > 1$ } \\
\sqrt{1 - \sqrt{\frac{\bigl(1 +
F(0)\bigr)^2}{(\beta\epsilon)\lambda^2a_2\left(N\over {\cal
S}\right) }}} & \mbox{;~${\frac{\bigl(1 +
F(0)\bigr)^2}{(\beta\epsilon)\lambda^2~a_2 {\frac{N}{\cal S}}}} < 1$ }
\end{array}
\label{equ44a}
\end{equation}
The above approximate solution is valid only in the vicinity of the
critical point $S \cong 0$, where the general solution for $\lambda$
in terms of $S$ can be approximated with its lowest order term. Away
from the critical point additional terms in the expansion of $\lambda
= \lambda (S)$ would have to be taken into account. The discussion in
this contribution will not be extended to this regime of orientational
order parameter.

\section{Discussion}

As already stated the existence of effective attraction between
segments of a polymer chain embedded in a flexible surface at non-zero
temperature is not completely unexpected. D'Hoker et al. \cite{bib8}
have found an effective (Casimir) attraction between beads embedded in
strings and membranes. Though their model applies to a different
situation than our own calculation, its consequences stemm
from the same basic physical mechanisms.

The effective nematic interaction of the form $ \int_0^N\!\!\int_0^N
dn dn' \dot{\brho}_i(n) \dot{\brho}_k(n) \\ \dot{\brho}_l(n')\dot{\brho}_m(n')
\left< {\bf
n}_i(n){\bf n}_m(n') \right> \left< {\bf n}_k(n'){\bf n}_l(n) \right>$
can in fact be viewed as a special type of Casimir force between
different segments of  a polymer chain. It is due to the change in the
energy of conformational fluctuations of the supporting surface (
'zero point energy' \cite{bib10} in the case of electromagnetic
Casimir forces) in the presence of embedded chains. It is analogous to
the zero order van der Waals interaction energy that is due purely to
thermodynamic fluctuations \cite{bib11}.

The dependence of this Casimir force on the orientation of the polymer
segments comes essentially from the geometric constraint for the
polymer chain, i.e. $\dot{\bf r}^2(n) = 1$, since some of the
3-D configurations are excluded if the chain is confined to lye in the
embedding 2-D surface. Its spatial dependence on the other hand, thus the
form of $F(\vert\brho\vert)$, stemms purely from the nature of the
bare surface fluctuations.

Once the presence of a nematic Casimir interaction is derived it
naturally leads to an orientational ordering transition. Also the
order of the transition clearly follows from the fact that the
effective polymer partition function corresponds to a polymer confined
to a 2-D surface. Nematic transition on a 2-D surface is of a second
order \cite{bib12}. Whether the order of the transition is preserved
for non-ideal polymers and/or higher orders in the perturbation
expansion remains to be assesed.

The present calculation is of importance for phenomena involving
interface embedded polymers. Monomolecular films of poly(dimethyl
siloxane) \cite{bib2} show domains of differing surface density even
below the submonolayer concentrations.  Orientational ordering
described in this contribution could possibly lead to domains of
different surface density of polymers at quite small average surface
densitites. One way of testing this mechanism is to quench the free
surface fluctuations and thus attenuate the nematic Casimir attraction
which should prevent the ordering of the embedded polymers.

Some of the open questions that we leave for further discussion is
whether the nature of the transition is conserved also in higher orders
of the perturbation expansion in terms of $F$ and what is the interplay
between the repulsive (steric) interactions and the nematic Casimir
attraction.



\vfill
\newpage
\begin{thebibliography}{99}

\bibitem{bib1} Ringsdorf, H., Schlarb, B., Venzmer, J.
{\sl Angew.Chemie} {\bf 100} (1988) 117.
\bibitem{bib2} Mann, E.K., H\' enon, S., Langevin, D., Meunier, J.,
{\sl J.Phys. II France} {\bf 2} (1992) 1683.
\bibitem{bib3} Dvolaitzky, M., Guedeau - Boudeville, M.A., Leger, L.,
{\sl Langmuir} {\bf 8} (1992) 2595.
\bibitem{bib4} Kozlov, M.M., Helfrich, W., {\sl Langmuir} {\bf 10} (1994)
4219 (and references therein).
\bibitem{bib5} Kozlov, M.M., Helfrich, W., {\sl Langmuir} {\bf 9}
(1993) 2761.\\
Kozlov, M.M., Helfrich, W., {\sl J.Phys. II France} {\bf 4}
(1994) 1427.\\
Kozlov, M.M., Helfrich, W., {\sl Phys.Rev. ~E} {\bf 51}
(1995) (in print).
\bibitem{kardar} Li, H., Kardar, M., {\sl Phys.Rev.} {\bf A 46} (1992) 6490.
\bibitem{bib6} Kardar, M., Orland, H., (1995) (cond-mat preprint)
\bibitem{bib7} Dalley, S., (1995) (cond-mat preprint)
\bibitem{bib8} D'Hoker, E., Sikivie, P., Kanev, Y., (1995) (cond-mat
preprint)
\bibitem{glob} Bawendi, M.G., Freed, K.F., {\sl J.Chem.Phys.} {\bf 83}
(1985) 2491. \\
Lagowski, J.B., Noolandi, J., {\sl J.Chem.Phys.} {\bf 95}(1991) 1266.
\bibitem{doi} Doi, M. and Edwards, S.F., {\sl The Theory of Polymer
Dynamics} (Clarendon Press, Oxford, 1986).
\bibitem{bib9} Gupta, A.M., Edwards, S.F., {\sl J.Chem.Phys.} {\bf 98}
(1993) 1588. \\
Gupta, A.M., Edwards, S.F., {\sl Polymer} {\bf 34} (1993) 3112.
\bibitem{bib10} Milloni, P.W., {\sl The Quantum Vacuum} (Academic
Press, New York, 1994).
\bibitem{bib11} Mahanty, J., Ninham, B.W., {\sl Dispersion forces}
(Academic Press, New York, 1976).
\bibitem{bib12} Frenkel, D., {\sl Statistical mechanics of liquid
crystals} in {\sl Liquids, Freezing and Glass Transition}, Eds.
Hansen, J.P., Levesque, D., Zinn-Justin, J. (Elsevier, Amsterdam, 1991).
\end{thebibliography}
\end{document}